\documentclass[lettersize,journal]{IEEEtran}
\usepackage{amsmath,amsfonts}
\usepackage{algorithmic}
\usepackage{algorithm}
\usepackage{array}
\usepackage[caption=false,font=normalsize,labelfont=sf,textfont=sf]{subfig}
\usepackage{textcomp}
\usepackage{stfloats}
\usepackage{url}
\usepackage{verbatim}
\usepackage{graphicx}
\usepackage{cite}
\usepackage{booktabs}
\usepackage{makecell}
\usepackage{stfloats}
\usepackage{float}
\usepackage{amsmath}
\usepackage{setspace}
\usepackage{MnSymbol,bbding,pifont}
\usepackage{lipsum,mwe,cuted}
\usepackage{arydshln}
\usepackage{fontawesome}
\begin{document}
{
\title{A Fully Self-Synchronized Control for Hybrid Series-Parallel Electronized Power Networks

\author{Zexiong Wei,
		Yao Sun,
	Xiaochao Hou,
	Mei Su
}
}

\markboth{}%
{Shell \MakeLowercase{\textit{et al.}}: A Sample Article Using IEEEtran.cls for IEEE Journals}

\IEEEpubid{}

\maketitle

\begin{abstract}
The hybrid series-parallel system is the final form of the power electronics-enabled power system, which combines the advantages of both series and parallel connections.
Although self-synchronization of parallel-type and series-type systems is well known, self-synchronization of hybrid systems remains unrevealed.
To fill in this gap, a fully self-synchronized control for hybrid series-parallel system is proposed in this paper.  
Based on the self-synchronization mechanism of power angle in parallel-type system and power factor angle in series-type system, a decentralized control strategy by integration of power droop and power factor angle droop can realize self-synchronization and power balancing of each module in the hybrid system.

\end{abstract}


\section{Introduction}
\IEEEPARstart{W}{ith}
the increasing development of renewable energy,  represented by photovoltaic and wind power\cite{blaabjergDistributedPowerGenerationSystems2017}, 
power electronics-enabled power systems are gradually replacing traditional generator-based power systems in certain locations\cite{strunzEnabling100Renewable2023}. 
Benefit from the flexibility and controllability of power electronics, power electronics-enabled power systems in various ways to connect, can be divided into parallel-type\cite{johnsonSynchronizationParallelSinglePhase2014b},\cite{wangLifetimeOrientedDroopControl2019}, series-type\cite{ouImprovedOphDroop2024a} and hybrid series-parallel\cite{geIntegratedSeriesParallelMicrogrid2018} systems as shown in Fig. \ref{fig_1}(a)-(c).
The hybrid system combines the advantages of both series and parallel connections, where low-voltage power sources are connected in cascade to form high-voltage strings, which are then connected in parallel to increase the power rating and redundancy of the system.
Nevertheless, the large number of inverter modules in such systems introduces control issues, such as synchronization and power sharing.

In order to address these control issues, the decentralized control methods have been widely studied due to the communication-free solution\cite{liuSynchronizationMethodModular2020}. 
In the parallel-type system, the most well-known method is classic droop control strategy\cite{chandorkarControlParallelConnected1993b}, which achieves self-synchronization among the parallel inverters.
Different from the parallel-type system, all modules in the series-type system share a common current, and the voltage sharing is equivalent to the power sharing.
In \cite{jinweiheInversePowerFactor2017a}, a power factor-frequency ($\cos\varphi-f$) inverse droop control strategy is proposed to first self-synchronize all modules in the islanded operation.
Nevertheless, this method is only available for resistive-inductive (RL) load.
To address this constraint, a power factor angle droop control is proposed in \cite{yaosunPowerFactorAngle2021a}, which presents the critical role of the power factor angle in the decentralized control of the series-type system.

Although the decentralized control method has distinct advantages and the hybrid series-parallel is very promising topology for high power application, to the best of our knowledge, there have been no previous literature about the hybrid system in the islanded operation with communication-free control strategy.
To fill in this gap, the decentralized control strategy belonging to the hybrid system is proposed.

 \begin{figure*}[!t]
 	\centering
 	\includegraphics[width=7.1in]{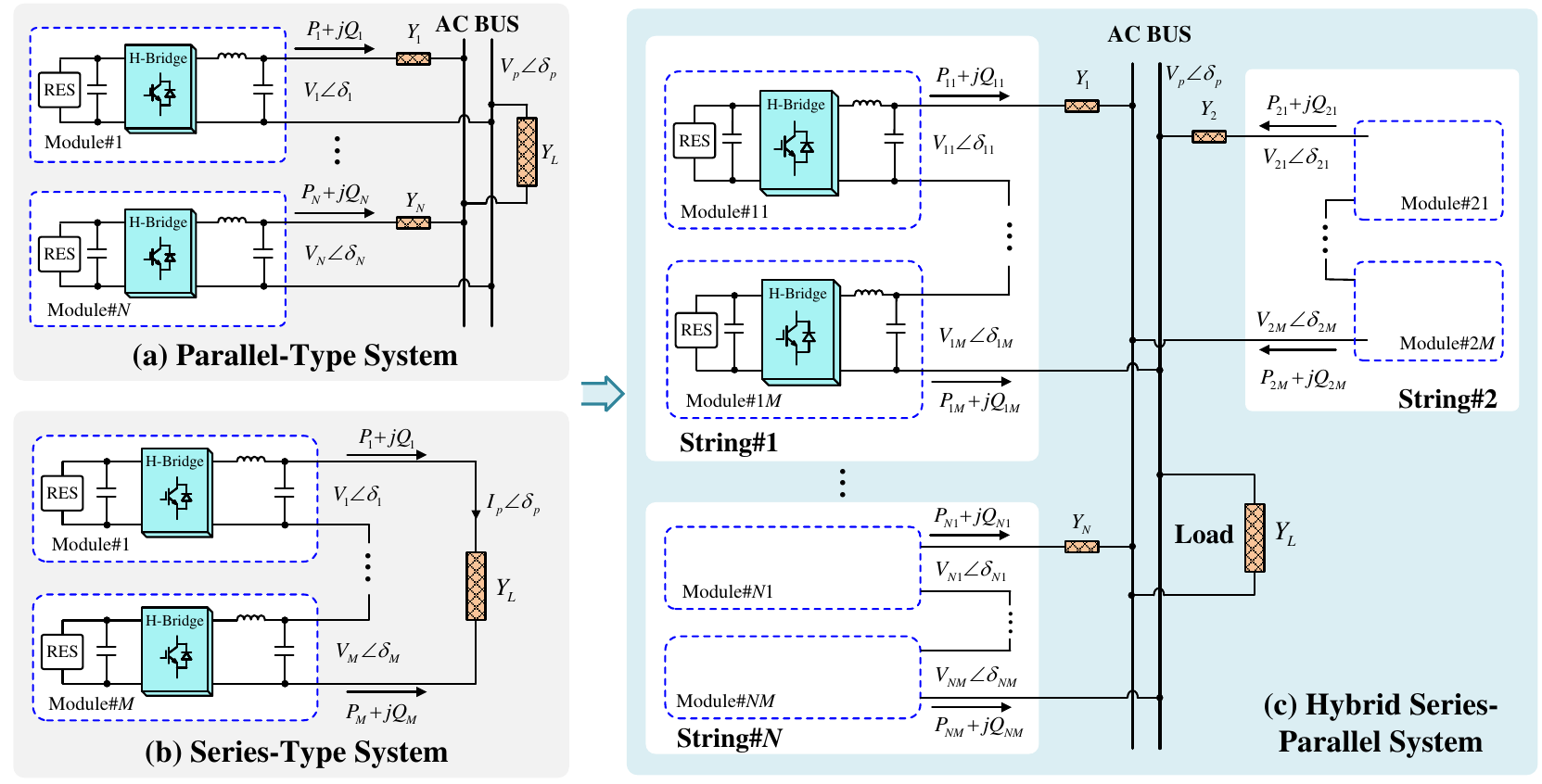}
 	  \vspace{-0.7cm}
 	\caption{The structures of power electronics-enabled power systems
 		(a) Parallel-type system. 
 		(b) Series-type system. 
 		(c) Hybrid series-parallel system.}
 	\vspace{-0.3cm}
 	\label{fig_1}
 \end{figure*}


\section{Decentralized Control of Hybrid Parallel-Type and Series-Type Systems}

\subsection{Power Characteristic of Hybrid Series-Parallel System}
The hybrid series-parallel system is topologically the combination of parallel-type and series-type systems.
The configuration of the hybrid series-parallel system is shown in Fig. \ref{fig_1}(c). 
There are a total of \textit{N} parallel strings, each consisting of $M$  modular inverters in series.

The output active power ${P_{ij}}$ and reactive power ${Q_{ij}}$ of the \textit{ij}-th modular inverter can be expressed as
\begin{equation}
	{P_{ij}} + j{Q_{ij}} = {V_{ij}}{e^{j{\theta _{ij}}}}{\left( {\left( {\sum\limits_{b = 1}^M {{V_{ib}}{e^{j{\theta _{ib}}}}}  - {V_P}{e^{j{\theta _P}}}} \right)\left| {{Y_i}} \right|{e^{j{\phi _i}}}} \right)^*}
	\label{eq2_1}
\end{equation}
where ${\delta _{ij}}$ and ${V_{ij}}$ are the phase angle and output voltage amplitude of the \textit{ij}-th modular inverter. ${\delta _{p}}$ and ${V_{p}}$ represent the phase angle and voltage amplitude of the AC bus. $\left| {{Y_i}} \right|$ and ${\phi _i}$ are the amplitude and angle of the equivalent line admittance in String\#\textit{i}. The voltage of the AC bus is calculated by Kirchhoff's principles as (\ref{eq2_2}).
\begin{equation}
	{V_P}{e^{j{\theta _P}}} = \sum\limits_{a = 1}^N {\sum\limits_{b = 1}^M {\left| {Y_{a}^{'}} \right|{V_{ab}}{e^{j{\theta _{ab}}{\rm{ + }}{{\phi}_{a}^{'}}}}} } 
	\label{eq2_2}
\end{equation}
where
\begin{equation}
	\begin{array}{l}
		Y_{a}^{'} = \displaystyle\frac{{{Y_a}}}{{\displaystyle\sum\limits_{c = 1}^N{{Y_c}} }+{Y_L}} = \left| {{{Y}_{a}^{'}}} \right|{e^{j\phi _{a}^{'}}}
	\end{array}
	\label{eq2_3}
\end{equation}
where ${Y_L}$ is load admittance.

By breaking down (\ref{eq2_1}) into real and imaginary components, the power transmission characteristic is given by
\begin{equation}
	\begin{array}{l}
		\displaystyle{P_{ij}} = {V_{ij}}\left| {{Y_i}} \right|\sum\limits_{b = 1}^M {{V_{ib}}\cos \left( {{\varphi _{ij}} - {\varphi _{ib}} - {\phi _i}} \right)} \\\displaystyle
		\;\;\;\;\quad- {V_{ij}}\left| {{Y_i}} \right|\sum\limits_{a = 1}^N {\sum\limits_{b = 1}^M {\left| {{Y_{a}^{'}}} \right|{V_{ab}}\cos \left( {{\delta _{ij}} - {\delta _{ab}} - {{\phi_{a}^{'}}} - {\phi _i}} \right)} } 
	\end{array}
	\label{eq2_4}
\end{equation}
\begin{equation}
	\begin{array}{l}
		\displaystyle{Q_{ij}} = {V_{ij}}\left| {{Y_i}} \right|\sum\limits_{b = 1}^M {{V_{ib}}\sin \left( {{\varphi _{ij}} - {\varphi _{ib}} - {\phi _i}} \right)} \\\displaystyle
		\;\;\;\;\quad- {V_{ij}}\left| {{Y_i}} \right|\sum\limits_{a = 1}^N {\sum\limits_{b = 1}^M {\left| {{Y_{a}^{'}}} \right|{V_{ab}}\sin \left( {{\delta _{ij}} - {\delta _{ab}} - {{\phi_{a}^{'}}} - {\phi _i}} \right)} } 
	\end{array}
	\label{eq2_5}
\end{equation}

\subsection{Proposed Control Strategy of the Hybrid System}

A decentralized control strategy is proposed for the hybrid system by combining $P - \omega $ droop control and $\varphi - \omega $ droop control, where $P - \omega $ droop control is responsible for the synchronization of the parallel modules and $\varphi - \omega $ droop control is responsible for the synchronization of the series modules.
The expression for the proposed control strategy of the single modular inverter are
\begin{equation}
{\omega _{ij}} = {\omega ^*} - mP_{ij}^{}- {k_\varphi }{\varphi _{ij}}
	\label{eq2_6}
\end{equation}
\begin{equation}
V_{ij}^{} = {V_{ref}}
	\label{eq2_7}
\end{equation}
where ${\omega _{ij}}$ and ${\varphi _{ij}}$ are the angular frequency and power factor angle of the \textit{ij}-th modular inverter, respectively. ${\omega ^*}$ is the rated angular frequency. ${m}$ and ${k_\varphi }$ are the droop coefficients. 
$V_{ref}$ is the voltage reference.

The control diagram is shown in Fig. \ref{fig5}.
It is worth noting that the proposed control strategy only needs the local information of the module, thus this strategy is a decentralized approach.

\begin{figure}[!t]
	\centering
	\includegraphics[width=3.2in]{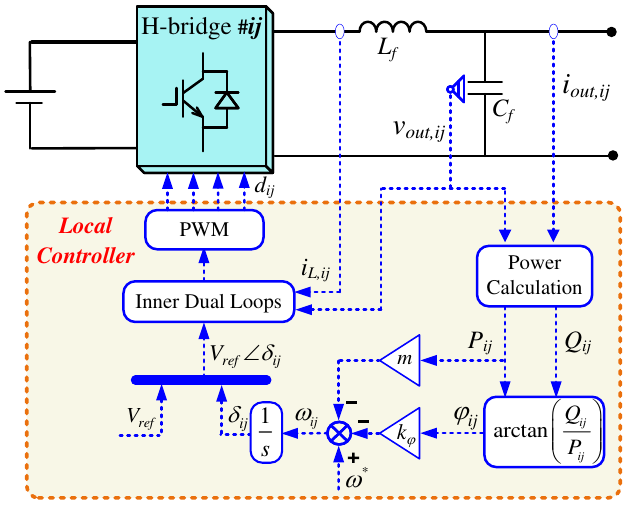}
	\caption{Control diagram of the \textit{ij}-th H-bridge inverter module.}
	\label{fig5}
\end{figure}

\subsection{Steady-State Analysis}

For the sake of simplicity in the subsequent analysis, it is assumed that the line admittance is the same for each string.
\begin{equation}
{Y_1} = {Y_2} =  \cdots  = {Y_N} = \left| {{Y_{{\rm{line}}}}} \right|{e^{j{\phi _{{\rm{line}}}}}}
	\label{eq2_0}
\end{equation}

(\ref{eq2_3}) is rewritten as
\begin{equation}
{Y'_a} = {Y_{eq}} = \left| {{Y_{eq}}} \right|{e^{j{\phi _{eq}}}}
	\label{eq3_1}
\end{equation}
%


Combining (\ref{eq2_4})-(\ref{eq3_1}), the power characteristics of the hybrid system are simplified as (\ref{eq2_111}) and (\ref{eq2_112}).
And the  power factor angle of the \textit{ij}-th module ($\varphi _{ij}$) is given as (\ref{eq3_3}).

\begin{figure*}[hb] 
	\centering 
	\hrulefill 
	\vspace*{8pt} 
	\begin{equation} 
		\begin{split}
{P_{ij}} = V_{ref}^2\left| {{Y_{{\rm{line}}}}} \right|\left( {\sum\limits_{b = 1}^M {\cos \left( {{\delta _{ij}} - {\delta _{ib}} - {\phi _{{\rm{line}}}}} \right)}  - \sum\limits_{a = 1}^N {\sum\limits_{b = 1}^M {\left| {{Y_{eq}}} \right|\cos \left( {{\delta _{ij}} - {\delta _{ab}} - {\phi _{eq}} - {\phi _{{\rm{line}}}}} \right)} } } \right)
		\end{split}
		\label{eq2_111}
	\end{equation}
		\begin{equation} 
		\begin{split}
{Q_{ij}} = V_{ref}^2\left| {{Y_{{\rm{line}}}}} \right|\left( {\sum\limits_{b = 1}^M {\sin \left( {{\delta _{ij}} - {\delta _{ib}} - {\phi _{{\rm{line}}}}} \right)}  - \sum\limits_{a = 1}^N {\sum\limits_{b = 1}^M {\left| {{Y_{eq}}} \right|\sin \left( {{\delta _{ij}} - {\delta _{ab}} - {\phi _{eq}} - {\phi _{{\rm{line}}}}} \right)} } } \right)
		\end{split}
		\label{eq2_112}
	\end{equation}
\end{figure*}

\begin{figure*}[hb] 
	\centering 
	\hrulefill 
	\vspace*{8pt} 
	\begin{equation} 
		\begin{split}
{\varphi _{ij}} = \displaystyle\arctan \frac{{{Q_{ij}}}}{{{P_{ij}}}}{\rm{ = }}\arctan \frac{{\displaystyle\sum\limits_{b = 1}^M \displaystyle{\sin \left( {{\delta _{ij}} - {\delta _{ib}} - {\phi _{{\rm{line}}}}} \right)}  - \displaystyle\sum\limits_{a = 1}^N {\sum\limits_{b = 1}^M {\left| {{Y_{eq}}} \right|\sin \left( {{\delta _{ij}} - {\delta _{ab}} - {\phi _{eq}} - {\phi _{{\rm{line}}}}} \right)} } }}{{\displaystyle\sum\limits_{b = 1}^M {\cos \left( {{\delta _{ij}} - {\delta _{ib}} - {\phi _{{\rm{line}}}}} \right)}  -\displaystyle \sum\limits_{a = 1}^N {\sum\limits_{b = 1}^M {\left| {{Y_{eq}}} \right|\cos \left( {{\delta _{ij}} - {\delta _{ab}} - {\phi _{eq}} - {\phi _{{\rm{line}}}}} \right)} } }}
		\end{split}
		\label{eq3_3}
	\end{equation}
\end{figure*}

In the steady-state, from (\ref{eq2_6}), we have
\begin{equation}
m{P_{11}}{\rm{ + }}{k_\varphi }{\varphi _{11}}{\rm{ = }} \cdots {\rm{ = }}m{P_{ij}}{\rm{ + }}{k_\varphi }{\varphi _{ij}}{\rm{ = }} \cdots {\rm{ = }}m{P_{NM}}{\rm{ + }}{k_\varphi }{\varphi _{NM}}
	\label{eq3_4}
\end{equation}

According to (\ref{eq2_111})-(\ref{eq3_4}), the same power angle of each module is an equilibrium point of the hybrid system.
\begin{equation}
{\delta _{11}} =  \cdots  = {\delta _{ij}} =  \cdots  = {\delta _{NM}}
	\label{eq2_9}
\end{equation}

\subsection{Small Signal Stability Analysis}
To analyze the stability of the proposed control strategy, the small signal analysis near the equilibrium point (\ref{eq2_9}) is carried out.
Assume that ${\delta _s}$ is the synchronous phase-angle of modular inverters in the steady state, and denote ${\tilde \delta _{ij}} = {\delta _{ij}} - {\delta _s}$. 
Since ${\dot{{\delta}}}_{ij} = {\omega _{ij}}$, by combining (\ref{eq2_6}) and (\ref{eq2_111})-(\ref{eq3_3}), and linearizing them around the equilibrium point (\ref{eq2_9}), we have
\begin{equation}
{\dot{{\tilde\delta}}}_{ij} =  - m{\tilde P_{ij}}- {k_\varphi }{\tilde\varphi _{ij}}
	\label{eq2_11}
\end{equation}
where ${{{\tilde P}_{ij}}}$ and $\tilde\varphi _{ij}$ are the small signals of active power and power factor angle, respectively.

Rewrite (\ref{eq2_11}) in matrix form as
\begin{equation}
\mathop {{\boldsymbol{\tilde \delta }}}\limits^{\bf{.}}  = {\bf{A}}{\boldsymbol{\tilde \delta }} =  \left( {m{{\bf{A}}_p} + {k_\varphi }{{\bf{A}}_\varphi }} \right){\boldsymbol{\tilde \delta }}
	\label{eq2_12}
\end{equation}
\vspace{0.5ex}where ${\boldsymbol{\tilde \delta }} = {\left[ {{{\tilde \delta }_{11}}, \cdots,{{\tilde \delta }_{NM}}} \right]^T}$,  ${{\bf{A}}_p} = {{\eta _{p1}}{\bf{L}}_{N \times M}^1 + {\eta _{p2}}{\bf{L}}_{N \times M}^2}$ and  ${{\bf{A}}_\varphi} =  {{\eta _{\varphi1}}{\bf{L}}_{N \times M}^1 + {\eta _{\varphi2}}{\bf{L}}_{N \times M}^2}$.
${\eta _{p1}}$, ${\eta _{p2}}$, ${\eta _{\varphi1}}$ and ${\eta _{\varphi2}}$ are given:
\begin{equation}
\left\{ \begin{array}{l}
	\vspace{1ex}
	\displaystyle{\eta _{p1}} = V_{ref}^2\left| {{Y_{{\rm{line}}}}} \right|\left| {{Y_{eq}}} \right|\sin \left( {{\phi _{eq}} + {\phi _{{\rm{line}}}}} \right)\\
	\vspace{1ex}
	\displaystyle{\eta _{p2}} =  - V_{ref}^2\left| {{Y_{{\rm{line}}}}} \right|\sin \left( {{\phi _{{\rm{line}}}}} \right)\\
	\vspace{1ex}
	\displaystyle{\eta _{\varphi 1}} = \frac{{\left( {\cos \left( {{\phi _{eq}}} \right) - N\left| {{Y_{eq}}} \right|} \right)\left| {{Y_{eq}}} \right|}}{{M + {N^2}M{{\left| {{Y_{eq}}} \right|}^2} - 2NM\left| {{Y_{eq}}} \right|\cos \left( {{\phi _{eq}}} \right)}}\\

	\displaystyle{\eta _{\varphi 2}} = \frac{{N\left| {{Y_{eq}}} \right|\cos \left( {{\phi _{eq}}} \right) - 1}}{{M + {N^2}M{{\left| {{Y_{eq}}} \right|}^2} - 2NM\left| {{Y_{eq}}} \right|\cos \left( {{\phi _{eq}}} \right)}}
\end{array} \right.
	\nonumber
	\label{eq2_0-1}
\end{equation}
where both ${{\bf{L}}_{N \times M}^1}$ and ${\bf{L}}_{N \times M}^2$ are Laplacian matrices, although they each describe different graphs.${\bf{L}}_{N \times M}^1 = \left( {N \times M} \right){{\bf{E}}_{N \times M}} - {{\bf{1}}_{N \times M}}$ and ${{\bf{L}}_M} = M{{\bf{E}}_M} - {{\bf{1}}_M}$. ${\bf{E}}{\rm{ = }}diag\left( {1, \cdots ,1} \right)$, ${\bf{1}}$ is an all-one matrix. The subscript indicates the order of the matrix.

The eigenvalues of system matrix ${\bf{A  }}$ are given as
\begin{equation}
\left\{ \begin{array}{l}
		\vspace{1ex}
	{\lambda _1}\left( {\bf{A}} \right){\bf{ = }}0\\
		\vspace{1ex}
	 {\lambda _2}\left( {\bf{A}} \right){\bf{ = }} \cdots  = {\lambda _N}\left( {\bf{A}} \right) = MN{\eta _1}\\

	{\lambda _{N + 1}}\left( {\bf{A}} \right){\bf{ = }} \cdots  = {\lambda _{NM}}\left( {\bf{A}} \right) = M\left( {N{\eta _1} + {\eta _2}} \right)
\end{array} \right.
	\label{eq3_7}
\end{equation}
where ${\eta _1}=m{\eta _{p1}}+{k_\varphi }{\eta _{\varphi1}}$ and ${\eta _2}=m{\eta _{p2}}+{k_\varphi }{\eta _{\varphi2}}$ are constants.

Therefore, the necessary and sufficient condition of system stability is expressed as
\begin{equation}
\left\{ \begin{array}{l}
		\vspace{1ex}
	{\lambda _p}=m{\eta _{p1}}+{k_\varphi }{\eta _{\varphi1}} < 0\\
	{\lambda _c}=m\left( {N{\eta _{p1}} + {\eta _{p2}}} \right) + {k_\varphi }\left( {N{\eta _{\varphi 1}} + {\eta _{\varphi 2}}} \right) < 0
\end{array} \right.
	\label{eq3_8}
\end{equation}
where ${\lambda _p}$ and ${\lambda _c}$ are denoted as eigenvalues of system matrix.

\section{Conclusion}
In conclusion, this paper proposes a fully self-synchronized control for hybrid series-parallel system.
The decentralized control strategy by integration of power droop and power factor angle droop can realize self-synchronization and power balancing of each module in the hybrid system. 
Moreover, the stability analysis is also carried out and the design range of the control parameters is given.
The paper provides the self-synchronization control framework for hybrid series-parallel system, which will promote the power electronics-enabled power system.

\bibliographystyle{IEEEtran}

\bibliography{IEEEabrv,ref.bib}

\vfill

\end{document}